\documentclass[12pt,A4]{article}
\usepackage{amsmath}
\usepackage{amsfonts}
\usepackage{graphicx,psfrag,epsf}
\usepackage{enumerate}
\usepackage{natbib}
\usepackage{apalike}
\usepackage{url} 
\usepackage{fullpage}
\usepackage{caption}
\usepackage{subfigure}
\newcommand{\bs}[1]{\boldsymbol{#1}}
\newcommand{\mc}[1]{\mathcal{#1}}
\newcommand{\bl}{} 
\newcommand{\el}{}
\newcommand{\pderiv}[2]{\frac{\partial #1}{\partial #2}}
\newcommand{\pderivsq}[1]{\frac{\partial^2}{(\partial #1)^2}}

\usepackage{natbib}

\title{\bf Sovereign Risk Indices and Bayesian Theory Averaging}
\author{Alex Lenkoski and Fredrik Lohne Aanes\\\emph{Norwegian Computing Center}}

\begin{document}
\linespread{1.15}
\maketitle
\thispagestyle{empty}
\begin{abstract}
  \noindent In economic applications, model averaging has found principal use examining the validity of various theories related to observed heterogeneity in outcomes such as growth, development, and trade.
  Though often easy to articulate, these theories are imperfectly captured quantitatively. A number of different proxies are often collected for a given theory and the uneven nature of this collection requires care when employing model averaging.  Furthermore, if valid, these 
  theories ought to be relevant outside of any single narrowly focused outcome equation.  We propose a methodology which treats theories as represented by latent indices, these latent processes controlled by model averaging on the proxy level.  To achieve generalizability of the theory index our framework assumes a collection of outcome equations.  We accommodate a flexible set of generalized additive models, enabling non-Gaussian outcomes to be included.  Furthermore, selection of relevant theories also occurs on the outcome level, allowing for theories to be differentially valid.  Our focus is on creating a set of theory-based indices directed at understanding a country's potential risk of macroeconomic collapse.  These Sovereign Risk Indices are calibrated across a set of different ``collapse'' criteria, including default on sovereign debt, heightened potential for high unemployment or inflation and dramatic swings in foreign exchange values.  The goal of this exercise is to render a portable set of country/year theory indices which can find more general use in the research community.
\end{abstract}
\linespread{1.5}

\section{Introduction}
In economic applications, Bayesian Model Averaging (BMA) has proven a useful tool to assess theories related to the potentials and risks of economic expansion, see \cite{steel2017model} for a comprehensive review.
All economic theories are in some sense qualitative and no single empirical observation can encapsulate the theory's essence perfectly.
To address this, a group of variables -- self-evidently correlated -- are often collected to proxy each theory.
Not accounting for the uneven manner by which different variables may be available for each theory can lead to inappropriate conclusions regarding overall theory validity.
Standard approaches to BMA can be modified, especially through the model prior, to account for these characteristics, but still consider the direct effect of the collected variables on the single response in question. One example is \cite{chen2017determinants}, which consider the determinants of the 2008 crisis. They use a hierarchical formulation that allows for a simultaneous selection of both theories and relevant variables.

We propose an entirely separate approach to testing theories, both with regards to standard BMA and also to \cite{chen2017determinants}, through a new model averaging approach. We assume each observation has a number of latent features encoding values for these theories. This requires the researcher to pre-specify which theory a given empirical observation is meant to proxy, a task which is often straightforward and frequently performed in practice.  The outcome of this modeling exercise is a set of theory indices associated with each observation, as well as the model parameters necessary to derive these indices for observations not included in training.  Our second innovation is to link the embedding of empirical factors to theory indices across a number of correlated outcome variables. This is driven by a motivation for theory index consistency.  Ideally, an index which assesses the strength of a government's institutions should be roughly the same when using the index to predict the potentials of economic growth and the susceptibility to economic collapse, for example.  Indeed an ideal encoding would allow the theory index to be trained on one set of outcome variables and be immediately useful as a standalone feature in modeling separate but related economic activity.
We therefore construct a framework by which theory-level modeling occurs on a latent level and is tuned to addressing a theory's role in explaining the variability of a number of economic outcome variables simultaneously.

\cite{brock2003policy} recommend considering both theory uncertainty (many theories can explain a phenomena) and variable (which empirical proxies should be used to explain each theory) uncertainty. Following this recommendation, model averaging in our Bayesian Theory Averaging (BTA) approach occurs on two separate levels. On the theory-level, a standard BMA formulation is used to determine which proxies for a given theory have the greatest relevance.
Our modeling is across multiple different outcome variables and a given theory may only be relevant for a subset of these outcomes.
Thus, we also perform theory averaging on the outcome level, allowing theories to selectively enter into each outcome under consideration.

Outcomes in economics can be quantified in a variety of manners and thus our framework is formulated to entertain a broader family of outcome sampling distributions than the Gaussian context to which most economic BMA applications have adhered \citep{steel2017model}. 
Indeed, our framework is organized to accommodate all generalized additive models (GAMs) (see e.g. \cite{hastie2017generalized} or \cite{wood2017generalized}) and quantile GAMS (qgams) \citep{fasiolo2017fast}.  Operationally, the posterior model space is explored via Markov Chain Monte Carlo (MCMC), see e.g. \cite{gamerman2006markov} or \cite{robert2013monte}, and model moves are efficiently performed via Conditional Bayes Factors (CBFs), \citep{karl2012instrumental}, which have been shown to be highly useful in related model averaging exercises \citep{lenkoski_2013,dyrrdal_et_2015}.

Our motivating example concerns developing useful theory-based indices for quantifying the potential for significant negative economic outcomes in macroeconomies, which we term Sovereign Risk Indices (SRIs).
These outcomes range across default on Sovereign debt,
the potential for high levels of inflation or unemployment,
and heightened risks instability in foreign exchange. Useful introductions to sovereign default are \cite{Roubini2005} and \cite{Savona2012}. Each of these outcomes have a number of theories which explain their variability.
These theories encapsulate institutional and financial characteristics of each country and overall aspects of the global economy at the time and are proxied by a large number of potential variables.  By modeling these outcomes jointly, we can construct a set of theory indices that are relevant for general research into macroeconomic extremes.  Our goal is to create a broad database of SRIs that can then be made available to the general research population, where each index has an clearly defined construction and encodes a well-articulated theory regarding economic well-being. Our data combines the data in \cite{Savona2012} with new data sources, as explained in Section~\ref{sec:data}. 

The structure of the article is as follows.  Section~\ref{sec:BTA} outlines BTA.  The specifics of the algorithm that performs posterior inference for BTA is rather involved and relegated to Appendix A.  Section~\ref{sec:SRI} contains our analysis of the data which constructs the SRIs while Section~\ref{sec:Conclude} concludes.

\section{Bayesian Theory Averaging}\label{sec:BTA}
Let $\bs{Y}_i$ be an $R$ dimensional response vector for observation $i$ and $\mc{D} = \{\bs{Y}_1, \dots, \bs{Y}_n\}$ be a collection of $n$ such observations.  Each variate $Y_{ir}$ in the vector $\bs{Y}_{i}$ is assumed to belong to a general field $\mc{F}_r$.  In this paper we consider examples where $\mc{F}_r$ is $\{0,1\}$, $\mathbb{R}$ and $\mathbb{R}_+$, though others such as $\mathbb{N}$, could easily be entertained.  We associate $Y_{ir}$ with an outcome distribution as
$$
Y_{ir} \sim g_{r}(\bs{\theta}_{ir})
$$
where $g_r$ is a general probability density or mass function and $\bs{\theta}_{ir}$ is a set of parameters assumed to have dimension $q_r$ which control $g_r$.  A given parameter $\theta_{irj}$, $j \in \{1, \dots, p_r\}$ is either assumed to be a global parameter and thus $\theta_{irj} = \theta_{krj}$ for all $i,k \in \{1, \dots, n\}$ or to have a linear form related to a $T$ dimensional set of indices $\bs{I}_{i} = (I_{i1},\dots, I_{iT})$ by the relationship
\begin{align*}
  \theta_{ir} = \alpha_r + \sum_{t = 1}^{T} \gamma_{rt} I_{it}.
\end{align*}
In this formulation $\gamma_{rt}$ can either be $0$ or $\gamma_{rt} \in \mathbb{R}$.  By convention if several $\gamma_{rt}$ are non-zero for a given index $t$ then one of these non-zero $\gamma_{rt}$ is set to $1$ to avoid issues related to identification. This matter is discussed at length in Appendix A.\\
\indent The variable $I_{it}$ is then referred to as the theory-$t$ index for observation $i$.  We further assume that the $I_{it}$ depends on a set of $p_t$ theory proxies $\bs{X}_{it}$ according to the linear model
\begin{align*}
  I_{it} &= \bs{X}_{it}'\bs{\beta}_t + \epsilon_{it}
\end{align*}
where $\epsilon_{it} \sim \mc{N}(0, \nu_t^{-1})$.  Due to issues of identification, the precision term $\nu_t^{-1}$, is considered fixed, see Appendix A for a more detailed discussion of this term and its role in balancing the system when theory-transitions are made.\\
\indent Associated with the parameter $\bs{\beta}_t$ is a model $M_t \subset \{1, \dots, p_t\}$ such that $\beta_{it} = 0$ when $i \not\in M_t$, a standard BMA formulation.  As the ``null'' model can be controlled by the $\gamma_{rt}$ parameter, we exclude $M_t = \emptyset$ from our consideration, see \cite{kourtellos_et_2019} for a motivation of this structure.  Writing $\bs{\beta}_{M_t}$ to represent the subvector of $\bs{\beta}_{t}$ not constrained to zero we assume
\begin{align*}
  \bs{\beta}_{M_t} &\sim \mc{N}(0, \nu_t \mathbb{I}_{p_{M_t}}),
\end{align*}
where $p_{M_t}$ is the size of model $M_t$.  Alternative priors for this model could have been considered, see our discussion in the Conclusions section. Again, the term $\nu_t$ adjusts the $\beta_{M_t}$ parameters and their associated prior to aid in identification matters when adjusting the theory indices $\bs{\gamma}_t$.\\
\indent Finally, the model $M_t$ can have a number of priors, see \citet{ley_steel_2009} for an overview of potential issues to consider when selecting this prior.  We follow \cite{kourtellos_et_2019} which builds on \cite{durlauf_et_2012} by choosing a model prior for which
$$
pr(M_t)\propto |C_{M_t}|
$$
where $C_{M_t}$ is the correlation matrix defined by the variables in $M_t$.  As in \cite{kourtellos_et_2019} we avoid of the awkward need to account for the null model encountered in \cite{durlauf_et_2012} by our restriction that $M_t \neq \emptyset$, handling this side case through the $\bs{\gamma}_t$ vector.\\
\indent The system outlined above then serves as the core latent process which drives the subsequent outcome variables.  Thus we see that the models $M_t$ investigate which proxies best encode a theory quantitatively while also accounting for the obvious model uncertainty in this formulation.  The $\bs{\gamma}_t$ term serves two purposes.  First, by examining its non-zero elements we see for which response equations a given theory is relevant.  Secondly, by requiring the first non-zero $\gamma_{rt}$ to be equal to $1$ and all others to be in $\mathbb{R}$ the $\bs{\gamma}_r$ term scales the latent indices to allow them to enter into model parameters differentially and indeed in opposite directions.\\
\indent Finally, the latent theory indices $I_{it}$ are potentially of greatest interest, as they are meant to encapsulate the way that the theory proxies affect the outcome equations of interest.  Again, as outlined in the Appendix, these terms suffer from potential identification issues when combined with the restrictions placed on a given $\bs{\gamma}_r$.  The $\nu_t$ term accommodates this under-identification and therefore, final interest focuses on the scale-free term $\tilde{I}_{it} = \nu_t I_{it}$.\\
\indent Choices for families $g_r$ that control the outcome variables are considerable.  In our application, we focus on three models.  The first is logistic regression.  In this case $Y_{ir} \in \{0,1\}$, $q_r = 1$ and
$$
pr(Y_{ir} = 1) = \frac{\exp(\theta_{i1})}{1 + \exp(\theta_{i1})}.
$$
In our applications we then assume that
\begin{equation}
\theta_{ir1} = \alpha_r + \sum_{t = 1}^{T}\gamma_{rt}I_{it}.\label{eq:mean_model}
\end{equation}
We use this logistic regression to model the probability that a country will default on its sovereign debt based on theory-indices.\\
\indent The second family considered corresponds to the non-central asymmetric Laplace variates.  In particular, $Y_{ir} \in \mathbb{R}, q_r = 2$ and we write
\begin{align*}
  pr(Y_{ir} | \theta_{ir1}, \theta_{ir2}, \tau) &= \tau (1 - \tau)\exp\left\{\theta_{ir2} - e^{\theta_{ir2}}\rho_{\tau}(Y_{ir} - \theta_{ir1})\right\}\\
  \rho_{\tau}(Y_{ir} - \theta_{ir1}) &= (Y_{ir} - \theta_{ir1})(\tau - \mathbf{1}\{Y_{ir} < \theta_{ir1}\}).
\end{align*}
In our application we assume that the log-precision parameter $\theta_{ir2}$ is constant across observations and thus write $\theta_{ir2} = \kappa_r$ while $\theta_{ir1}$ has the form in (\ref{eq:mean_model}).  This model is often referred to as the Bayesian Quantile Regression since its posterior mode is related to the quantile regression estimate under the so-called \emph{pin-ball} loss $\rho_{\tau}$.  We employ this model for two separate variates, the inflation and unemployment rates and set $\tau = .9$ for both, thus focusing on the 90Th percentile of the respective distributions.\\
\indent Finally, we consider the Generalized Extreme Value (GEV) model as parameterized by
\begin{align*}
pr(Y_{ir} | \theta_{ir1}, \theta_{ir2}, \theta_{ir3}) =  e^{\theta_{ir2}} h(Y_{ir})^{-(\theta_{ir3} + 1)/\theta_{ir3}} \exp \Big( - h(Y_{ir})^{-\theta_{ir3}^{-1}}\Big), 
\end{align*}
for $h(Y_{ir}) > 0$ with 
\begin{align*}
h(Y_{ir}) = 1 + \theta_{ir3} e^{\theta_{ir2}} (Y_{ir} - \theta_{ir1}).
\end{align*}
The GEV model is used to model block maxima and hence understand the nature of extreme behavior.  In our case we use it to model the largest daily percentage jump in a country's exchange rate (relative to USD) seen over the course of a year.  We currently fix both the log precision $\theta_{ir2}$ and shape $\theta_{ir3}$ parameters to be constant and write these as $\kappa$ and $\xi$ respectively.\\
\indent Based on $\mc{D}$ we then conduct posterior interference on the full parameter set, which includes global parameters in the $\bs{\theta}_r$, theory-level models $M_t$, theory-inclusion and scaling parameters $\gamma_{rt}$ and linear model parameters $\bs{\beta}_{rt}$. Posterior inference is performed via Markov Chain Monte Carlo (MCMC).  Given the involved and nested nature of the MCMC, several different approaches are employed at different stages of the hierarchy and the full details are provided in Appendix A.\\
\indent However, the main themes of the MCMC involve conditional Bayes Factors (CBFs) to change models $M_t$ and update proxy regression parameters $\bs{\beta}_{t}$.  Standard block Metropolis-Hastings proposals using local Laplacian calculations of the log posterior density are used to update latent theory indices $I_{rt}$ as well as any global parameters in $\bs{\theta}_r$.  Finally, reversible jump methods \citep{green_1995} alternate $\gamma_{rt}$ between being $0$ or in $\mathbb{R}$, with a modicum of book keeping to ensure that at least one $\gamma_{rt}$ is set to $1$ when theory $t$ is represented in more than one dependent equation $r$, again to ensure identification of the system.\\
\indent In general, mixing of the algorithm appears relatively satisfactory, especially when using CBFs to compare non-neighboring models $M_t$. Run-time is on the order of hours for our focus dataset, as opposed to days or weeks. More details are provided in Appendix A.\\

\section{Using BTA to construct Sovereign Risk Indices}\label{sec:SRI}
\subsection{Data Outline}\label{sec:data}
Our dataset for constructing SRIs is based on the data used in \cite{Savona2012}.  \cite{Savona2012} track $70$ countries between the years of 1975 and 2010 and are primarily focused on whether a country defaults on its sovereign debt in a given year.  To model this default probability, \cite{Savona2012} collect $27$ covariates.  These covariates are meant to proxy $5$ different theories related to sovereign debt default.  In particular, they entertain the concepts of Insolvency, Illiquidity, Macroeconomic Factors, Political Factors and Global Systemic factors.  Table~\ref{tab:variables} provides an overview of the $27$ covariates considered and to which theory \cite{Savona2012} associate them.   For a given year, most covariates are ``lagged''(except for contagion, dummy for oil and dummy for international capital markets), in that these values would be available at the start of a given year, as opposed to co-occuring with the default event.  In total, these data correspond to 1998 country/year pairs.  Covariate missingness is prevalent, we use the imputed values derived from the methodology outlined in \cite{Savona2012}.  In the conclusions section we discuss alternative approaches that could have been incorporated directly into our methodology to handle covariate missingness.\\
\linespread{1.1}
\begin{table}\caption{Overview of variables considered in SRI dataset}\label{tab:variables}
  \begin{tabular}{lll}
    \hline\hline
    Theory & Short Variable Name & Description\\
    \hline
    Insolvency & MAC &  Market access to capital markets, dummy \\
    Insolvency & IMF & IMF lending dummy\\
    Insolvency & CAY & Current account balance, in \% of GDP\\
    Insolvency & ResG & Reserves growth/change in \% \\
    Insolvency & XG & Export growth in \% \\
    Insolvency & WX & Export in USD billions \\
    Insolvency & TEDX & Total external debt to exports, in \% \\
    Insolvency & MG & Import growth, in \% \\
    Insolvency & FDIY & Foreign direct investment to GDP, in \% \\
    Insolvency & FDIG & Change in \% of foreign direct investment inflows\\
    Insolvency & TEDY & Total external debt to GDP, in \% \\
    Insolvency & SEDY &  Short term external debt to GDP, in \% \\
    Insolvency & PEDY & Public external debt to GDP, in \% \\
    Insolvency & OPEN & Exports and imports over GDP, in \% \\
    Illiquidity & STDR & Short term debt to reserves\\
    Illiquidity & M2R & M2 to reserves\\
    Illiquidity & DSER & Debt service on long term debt to reserves\\
    Macroeconomic & DOil & Oil producing dummy\\
    Macroeconomic & RGRWT & Real (inflation adjusted) GDP change in \%  \\
    Macroeconomic & OVER & Exchange rate residual over liner trend \\
    Macroeconomic & UST & US treasury bill\\
    Political & PR & Index of political rights, 1 (most free) to 7 (least free)\\
    Political & History & Number of past defaults \\
    Systemic & Cont\_tot & Number of defaults in the world \\
    Systemic & Cont\_area & Number of defaults in the region the country is part of\\
    \hline\hline
  \end{tabular}
\end{table}
\linespread{1.5}
\indent \cite{Savona2012} is concerned with predictive models of sovereign default and therefore solely focus on this binary outcome.  We augment the default binary with three other measures that can also indicate a macroeconomy in a state of collapse.  First, the country's lagged (i.e. one-year behind) inflation rate was originally included in the Macroeconomic factors group of covariates in \cite{Savona2012}.  We instead treat (non-lagged) inflation as another dependent variable and note that doing so has no effect on the Default outcome; a run of BTA solely on Default with inflation included in the Macroeconomic factors gave this variable 0 inclusion probability.  In addition, we collected unemployment data from the IMF website.  These data were only available for a subset (897 country pairs) of the total data.  We note that this dependent variable missingness poses no substantive problem in terms of the derivation of SRIs.  The BTA approach simply ignores the missing likelihood contributions necessary to update the asssociated latent theory indices.

\indent Finally, we collected foreign exhange rate data from the website of IMF.  For each country/year pair, we first computing the log rate change relative to the US dollar and then used the annual maximum of these log changes.  This variable therefore shows the largest single-day weakening of a currency relative to the US dollar in the course of a year, since one can buy more of the currency for the same amount of US dollars.  We avoided commercial sources of foreign exchange data and therefore only had these values for 272 country/year pairs.  See our discussion in the conclusions section regarding expanding these data.

\indent A country is in default if it is classified so by Standard \& Poor’s (SP) or if it receives a large nonconcessional loan by the IMF. A nonconcessional loan is a loan that has the standard IMF’s market-related interest rate, while a concessional loan has a lower interest rate. The type of loans we consider from IMF must be in excess of 100 percent of quota. Each member country has a quota, where the initial quota is set when a country joins IMF. The quota determines, by other things, the country’s access to IMF loans and for instance its voting power. By augmenting the data from SP with data from IMF, we also capture near-defaults or debt restructurings avoided through loan packages from IMF. We consider Stand By Agreements (SBA) and Extended Fund Facility (EFF).  

\indent Our posterior inference is performed after running the BTA algorithm for 400 thousand iterations over these data.  In order to verify convergence, 30 seperate runs of the algorithm were run simultaneously and the resulting output was inspected to verify posterior inference for each individual chain was nearly identical.  Runtime on a 32 core machine (dual 8-core 3.4 GHz AMD Ryzen processors with hyperthreading capabilities) with 128 GB of RAM was roughly 7.5 hours.
\subsection{Results}
\indent We begin our discussion of the SRI results by investigating outcomes on the theory level.  Table ~\ref{tab:theory_probs} shows the theory inclusion probability (i.e. the proportion of time that $\gamma_{rt}$ was not constrained to zero in the chain) for each theory, across the four outcome variables.  Given that the original dataset was constructed to model the default variable, it is unsurprising that all theories achieve inclusion probabilities of one for this outcome. The inflation outcome is interesting in that it suggests that proxies measuring a country's political stability have the best explanation for the upper tails of the inflation distribution.  All other theories are also relevant to inflation, achieving probabilities between .38 and .41, but nowhere near as strong as the political factors.\\
\indent The results for unemployment in Table~\ref{tab:theory_probs} suggests a more bimodal inclusion result.  The Insolvency and Systemic theories barely enter into the final outcome while the Illiquidity, Macroeconomic and Political factors achieve probabilities of one.  Finally, we achieve relatively low inclusion probabilities for all theories for the devaluation outcome.  This is likely due in part to the relatively small amount of data that was available using public sources, see our discussion in Section~\ref{sec:Conclude}.  However, we feel this result highlights a useful feature of BTA, namely that including this outcome variable and having the system set theory-inclusion probabilities to zero meant there was no subsequent effect on the calculation of theory indices.
\begin{table}[p]\caption{Theory Inclusion Probabilities by Dependent Variable}\label{tab:theory_probs}
  \begin{center}
    \begin{tabular}{lllll}
\hline\hline
  & Default & Inflation & Unemployment & Devaluation\\
\hline
Insolvency & 1 & 0.397 & 0.017 & 0.079\\
Illiquidity & 1 & 0.381 & 1 & 0.181\\
Macroeconomic & 1 & 0.383 & 1 & 0.086\\
Political & 1 & 0.991 & 1 & 0.061\\
Systemic & 1 & 0.419 & 0.015 & 0.071\\
\hline\hline
    \end{tabular}
    \end{center}
\end{table}

\indent Table ~\ref{tab:theory_means} shows the mean value (conditional on inclusion) of the parameter $\gamma_{rt}$ for each theory and outcome pair.  Since Default was ordered first in our system and achieves inclusion probabilities of $1$ for all theories, this system serves to orientate the rest of the outcomes.  In particular, a positive $\gamma_{rt}$ for the outcomes indicates that the directionality of this theory on the outcome is similar to that of default.  The conditional means on the inflation outcome reflect even more clearly the importance of the Political theory relative to all others.  The value its conditional mean achieves (1.905) is more than ten times the next highest conditional mean (.124 for the Macroeconomic factor).  Since the Political Theory achieved substantially higher inclusion probabilities in Table~\ref{tab:theory_probs} this implies that the unconditional effect of the political theory index is the main driver of the upper tail of inflation.\\
\indent Recalling again from Table~\ref{tab:theory_probs} that the unemployment outcome was driven by the Illiqudity, Macroeconomic and Political factors, the results in Table~\ref{tab:theory_means} are interesting.  They show that Illiquidity has a very strong, positive effect on the unemployment outcome.  We note that ``positive'' is in the sense of working in the same direction as Default.  The Macroeconomic and Political factors then balance this behavior; they are negatively orientated to the impact these factors have on Default, though of smaller magnitude to the Illiquidity theory.  Finally, as noted in Table~\ref{tab:theory_probs} there appears to be negligible effect of the theory indices on the devaluation outcome.
\begin{table}\caption{Mean value of $\gamma_{rt}$ conditional on inclusion}\label{tab:theory_means}
  \begin{center}
    \begin{tabular}{lllll}
\hline\hline
  & Default & Inflation & Unemployment & Devaluation\\
\hline
Insolvency & 1 & -0.053 & 0 & -0.004\\
Illiquidity & 1 & 0.061 & 2.32 & 0.018\\
Macroeconomic & 1 & 0.124 & -0.503 & -0.005\\
Political & 1 & 1.905 & -0.857 & 0\\
Systemic & 1 & -0.023 & 0 & 0.002\\
\hline\hline
    \end{tabular}
  \end{center}
\end{table}

\indent Table~\ref{tab:insolvency_results} shows the inclusion probabilities and conditional posterior mean for each proxy contained in the Insolvency theory group.   Five factors achieve probabilities of one.  These include two factors that measure the strength of the country's balance sheet, namely the country's current account (CAY) and reserves growth (ResG). In addition, two factors that measure a country's trade balance, namely total exports balance  and import growth (WX and MG respectively) are also included in the Insolvency theory with probability one along with a measure of the country's foreign obligations (public external debt to GDP, PEDY).  We see from Table~\ref{tab:insolvency_results} a set of proxies which together can give a view of the country's trade balance and its relation to the country's balance sheet, all in the context of external obligations.  Other variables such as the IMF lending dummy, a measure of Foreign Direct Investment (FDIY) and a more nuanced view of the country's debt load by breaking out short term debt (SEDY) also feature in the insolvency theory with inclusion probabilities of between 0.4 and 0.62.  The inclusion probability of all other proxies is roughly zero.
\begin{table}\caption{Proxy Level Results for the Insolvency Theory}\label{tab:insolvency_results}
  \begin{center}
    \begin{tabular}{lll}
\hline\hline
  & Probability & Conditional Mean\\
\hline
MAC & 0.063 & 0.024\\
IMF & 0.43 & -0.371\\
CAY & 1 & -0.066\\
ResG & 1 & -0.003\\
XG & 0.003 & 0\\
WX & 1 & -0.005\\
TEDX & 0.237 & 0\\
MG & 1 & -0.011\\
FDIY & 0.625 & -0.024\\
FDIG & 0 & 0\\
TEDY & 0.073 & -0.004\\
SEDY & 0.522 & 0.016\\
PEDY & 1 & -0.007\\
OPEN & 0 & 0\\
\hline\hline
    \end{tabular}
  \end{center}
\end{table}

\indent Table~\ref{tab:illiquidity_results} shows the inclusion probabilities for the Illiquidity theory.  In contrast to the balanced view offered in the Insolvency results of Table~\ref{tab:insolvency_results}, Table~\ref{tab:illiquidity_results} puts almost all weight on a single feature, a measure of a country's long-term debt service burden, namely DSER.  In some sense, we find this result appealing, as a more technical measure of money supply (M2R) cannot in itself indicate whether illiquidity events are likely to arise.  Furthermore, it is interesting to note that similar to the higher inclusion probability placed on long-term debt over short-term debt in Table~\ref{tab:insolvency_results} (i.e. PEDY versus SEDY) we see in the insolvency group that the short term debt to reserves (STDR) is given a weight of zero when included alongside the factor focused on longer term debt servicing (DSER).\\
\begin{table}\caption{Proxy Level Results for the Illiquidity Theory}\label{tab:illiquidity_results}
  \begin{center}
    \begin{tabular}{lll}
\hline\hline
  & Probability & Conditional Mean\\
\hline
STDR & 0.004 & 0\\
M2R & 0 & 0\\
DSER & 1 & 0.107\\
\hline\hline
    \end{tabular}
  \end{center}
\end{table}
\indent Table~\ref{tab:macro_results} shows the inclusion probabilities for the proxies in the Macroeconomic grouping.
We see that measures related to inflation dynamics (RGRWT) and exchange rate fluctuations (OVER) are given low inclusion probabilities.  Instead, a measurement of whether a country is dependent on Oil proceeds (DOil) and a technical factor related to global interest rates (UST) are the main two constituents of this theory with inclusion probabilities of .6 and 1 respectively.\\
\begin{table}\caption{Proxy Level Results for the Macroeconomic Theory}\label{tab:macro_results}
  \begin{center}
    \begin{tabular}{lll}
\hline\hline
      & Probability & Conditional Mean\\
\hline
DOil & 0.541 & 0.215\\
RGRWT & 0.009 & -0.005\\
OVER & 0.103 & 0\\
UST & 1 & 0.292\\
\hline\hline
    \end{tabular}
  \end{center}
\end{table}
Tables~\ref{tab:political_results} and~\ref{tab:systemic_results} show the inclusion probabilities for proxies of the Political and Systemic theories respectively.  In each theory there are only two features and all four receive high inclusion probabilities ranging form 0.7 to 1.  We see that the Political theory is thus a blend of the Political rights index (PR) and a measure of past susceptibility to default (History).  Likewise, a measure of global contagion (Cont\_tot) as well as local factors (Cont\_area) combine to form the Systemic theory, with slightly less weight (.76) placed on the local proxy.  
\begin{table}\caption{Proxy Level Results for the Political Theory}\label{tab:political_results}
  \begin{center}
    \begin{tabular}{lll}
\hline\hline
      & Probability & Conditional Mean\\
\hline
PR & 1 & 0.152\\
History & 1 & 1.102\\
\hline\hline
    \end{tabular}
  \end{center}
\end{table}
\begin{table}\caption{Proxy Level Results for the Systemic Theory}\label{tab:systemic_results}
  \begin{center}
    \begin{tabular}{lll}
\hline\hline
      & Probability & Conditional Mean\\
\hline
Cont\_tot & 0.995 & 0.04\\
Cont\_area & 0.76 & 0.081\\
\hline\hline
    \end{tabular}
  \end{center}
\end{table}
\indent We conclude by investigating detailed results for two of the theories, namely Insolvency and  Illiquidity.  Table~\ref{tab:insolvency_index} shows the country/year pairs with the five lowest and five highest posterior mean values of $I_{it}$ for the insolvency theory. The lowest five country/year pairs listed represent the countries whose Insolvency index indicates the lowest probabilities of default.  Interestingly, Gabon is represented twice amongst these five countries (for the years 1981 and 1995), which is unsurprising given the country's oil wealth and relative aggregate prosperity amongst African nations.  Amongst the five country/year pairs with the highest Insolvency index scores we see a mix of African (Tunisia 1988, Niger 1983) Caribbean (Trinidad and Tobago 1987, Haiti 1979) and Southeast Asian (Sri Lanka 2009) countries. Two of the five (i.e. 40\%) of these pairs experience a default, which is substantially higher than the 6\% average over all the data, showing the degree to which this feature is positively associated with default.\\
\begin{table}\caption{Highest and Lowest Five Country/Year Pairs for the Insolvency Theory}\label{tab:insolvency_index}
  \begin{center}
    \begin{tabular}{lllllll}
\hline\hline
  & Year & Insolvency & Default & Inflation & Unemployment & Devaluation\\
\hline
Gabon & 1995 & -13.683 & 0 & 36.116 &  NA  &  NA \\
Moldova & 1994 & -6.704 & 0 & 35.749 &  NA  &  NA \\
Korea, Rep. & 2009 & -6.157 & 0 & 4.704 & 3.2 & 0.071\\
Portugal & 1995 & -5.664 & 0 & 5.214 & 6.713 & 0.018\\
Gabon & 1981 & -5.543 & 0 & 12.34 &  NA  &  NA \\
Trinidad and Tobago & 1987 & 4.407 & 0 & 7.694 & 9.37 &  NA \\
Tunisia & 1988 & 4.437 & 1 & 8.226 &  NA  &  NA \\
Sri Lanka & 2009 & 4.448 & 0 & 22.564 & 5.22 & 0.012\\
Niger & 1983 & 4.703 & 1 & 11.642 &  NA  &  NA \\
Haiti & 1979 & 4.864 & 0 & -2.674 &  NA  &  NA \\
\hline\hline
    \end{tabular}
  \end{center}
\end{table}
\indent Table~\ref{tab:illiquid_index} shows the five highest and lowest country year pairs according to the illiquidity index.  Burundi in 1991 (i.e. two years before the start of the civil war that ran between 1993 and 2005) receives the lowest Illiqudity index, otherwise followed by countries in South Asia.  On the highest end, we see both Jamaica and Lesotho represented twice.  In addition, Gabon in 2002 is present, a year in which the country defaulted on its sovereign debt.  This contrast to Table~\ref{tab:insolvency_index} is illuminating, as it shows the trade off between potential for insolvency and risks of illiquidity in precipitating sovereign default.  We note again that two of the top five country year pairs record a default, similar to the results of Table~\ref{tab:insolvency_index}.  However, when inspecting the unemployment result, we also see high levels of unemployment for four of the five top countries (and a missing value for Gabon 2002, the remaining country). Simultaneously the countries with the lowest illiquidity indices have negligible unemployment rates. This lines up with the results presented in Table~\ref{tab:insolvency_results}, where the insolvency index had a large, positive effect on the unemployment outcome equation.
\begin{table}\caption{Highest and Lowest Five Country/Year Pairs for the Illiquidity Theory}\label{tab:illiquid_index}
  \begin{center}
    \begin{tabular}{lllllll}
\hline\hline
  & Year & Illiquidity & Default & Inflation & Unemployment & Devaluation\\
\hline
Burundi & 1991 & -5.391 & 0 & 7.002 & 0.48 &  NA \\
Pakistan & 1976 & -5.223 & 0 & 20.905 & 1.7 &  NA \\
Bangladesh & 1997 & -5.154 & 0 & 2.377 & 2.51 &  NA \\
Malaysia & 2007 & -5.079 & 0 & 3.609 & 3.32 & 0.012\\
Indonesia & 1977 & -5.07 & 0 & 19.859 & 1.92 &  NA \\
Jamaica & 1986 & 7.266 & 0 & 25.673 & 33.39 &  NA \\
Lesotho & 2009 & 7.311 & 0 & 10.721 & 35.46 &  NA \\
Lesotho & 1998 & 7.329 & 0 & -100 & 37.94 &  NA \\
Gabon & 2002 & 8.162 & 1 & 2.138 &  NA  &  NA \\
Jamaica & 1981 & 9.208 & 1 & 27.308 & 35.51 &  NA \\
\hline\hline
    \end{tabular}
  \end{center}
\end{table}

\begin{figure}
  \subfigure[Insolvency]{\includegraphics[width = .32\linewidth]{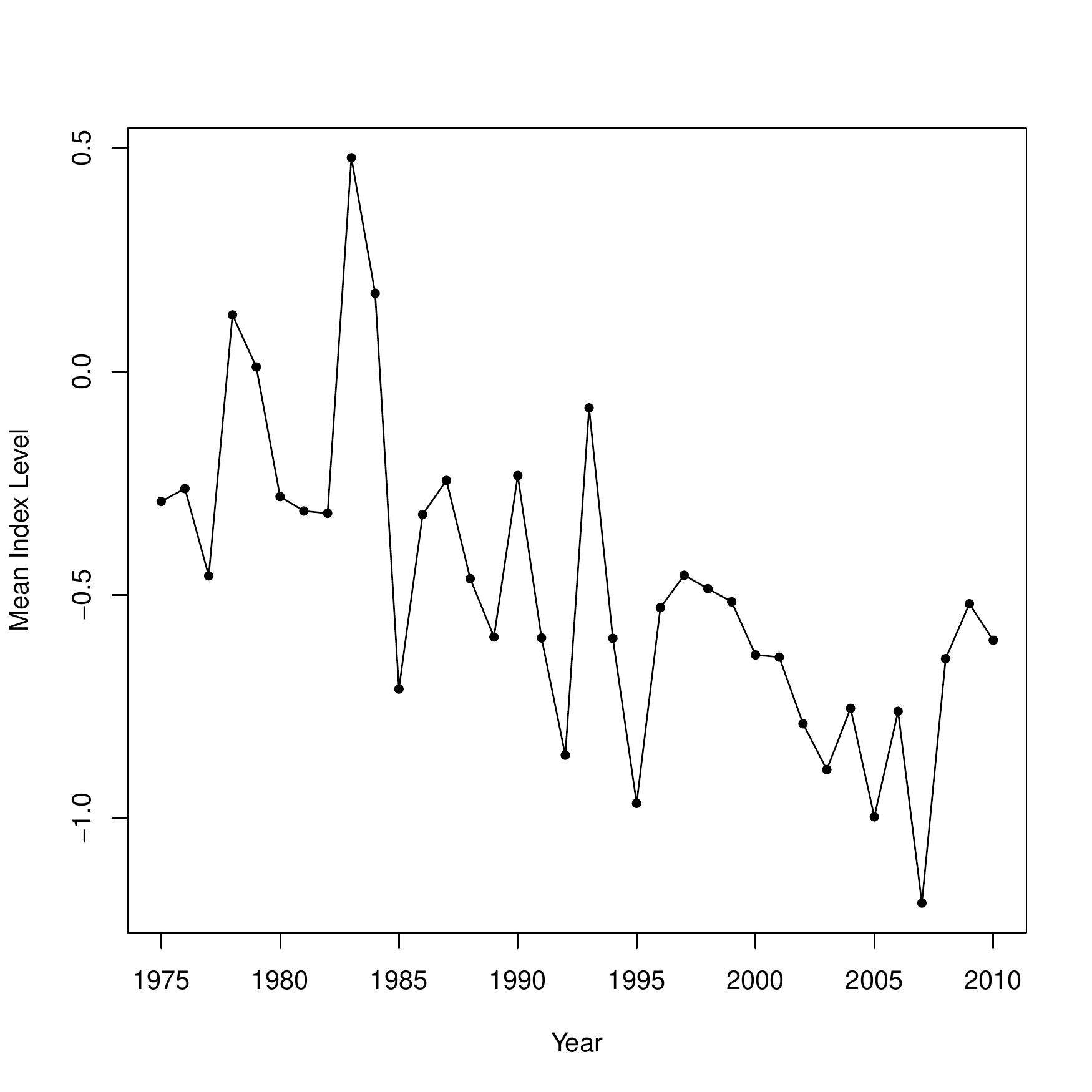}}
  \subfigure[Illiquidity]{\includegraphics[width = .32\linewidth]{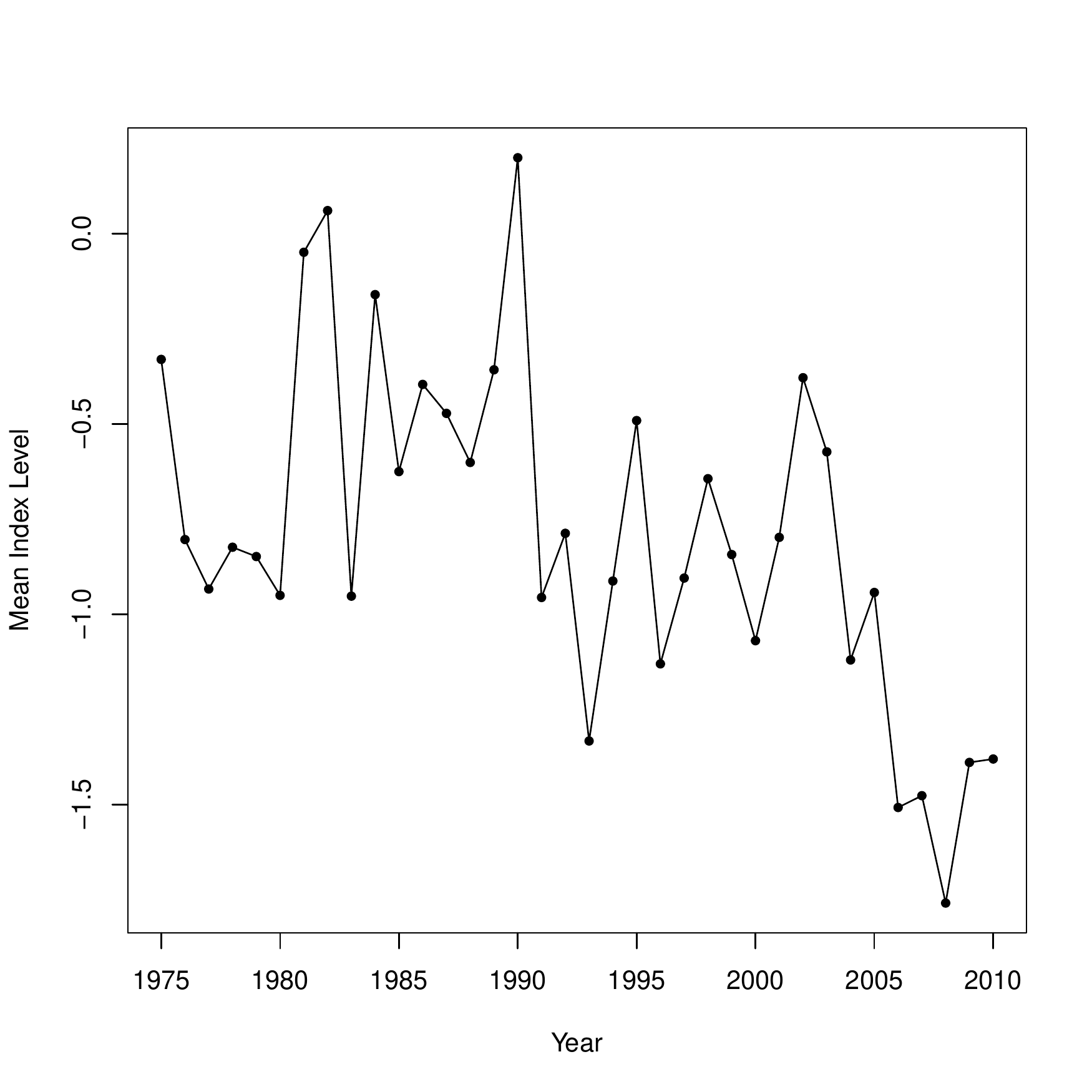}}
  \subfigure[Macroeconomic]{\includegraphics[width = .32\linewidth]{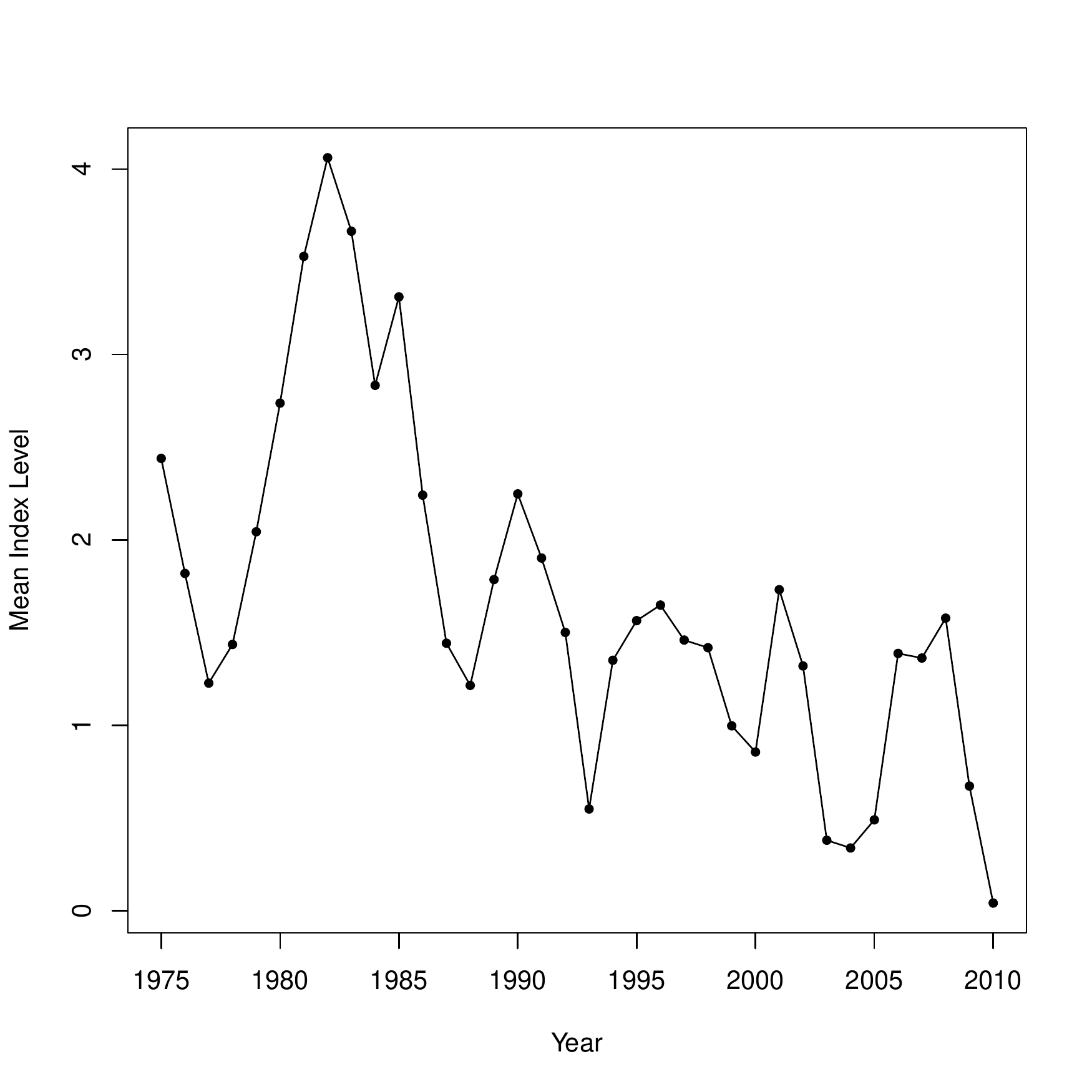}}
  \subfigure[Systemic]{\includegraphics[width = .32\linewidth]{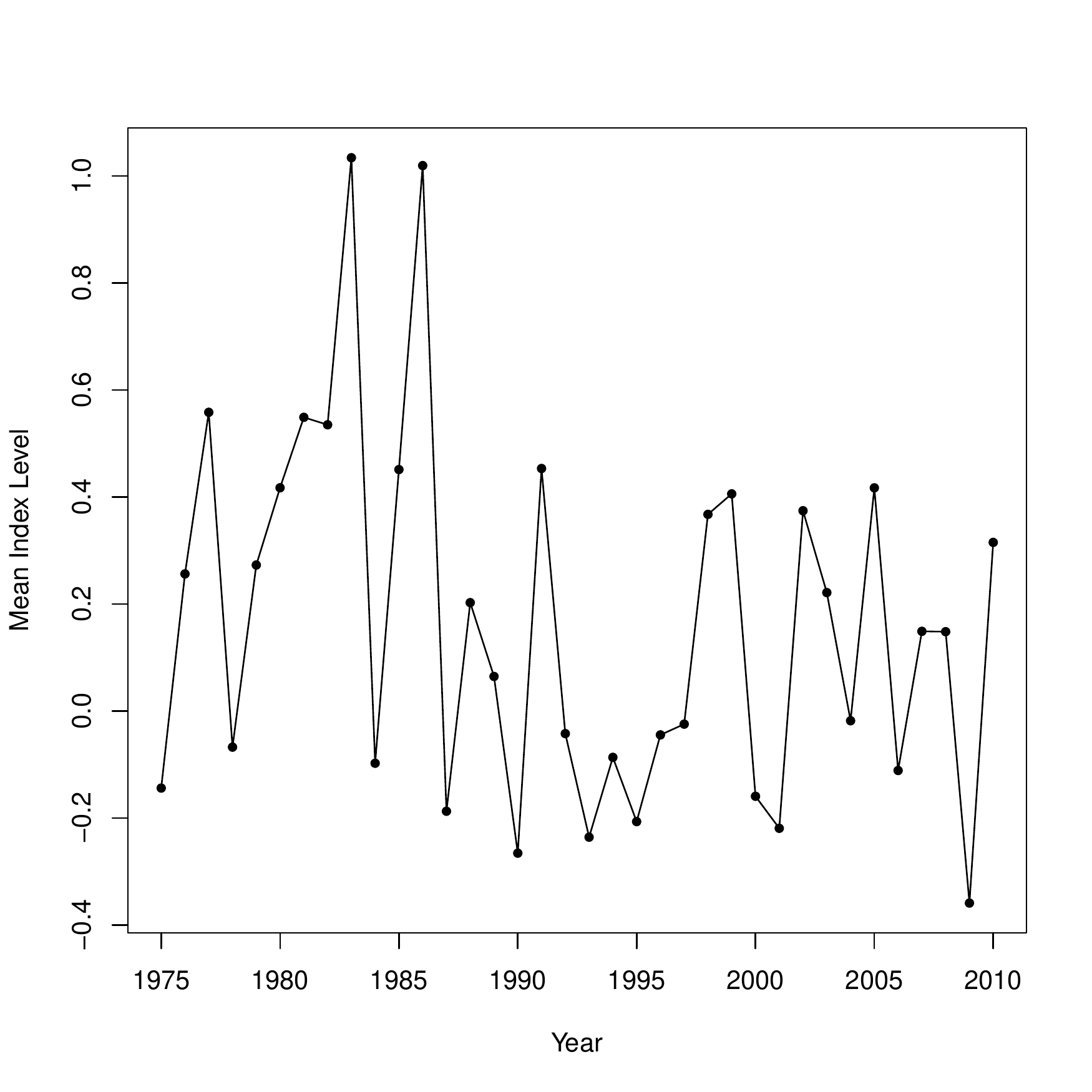}}
  \subfigure[Political]{\includegraphics[width = .32\linewidth]{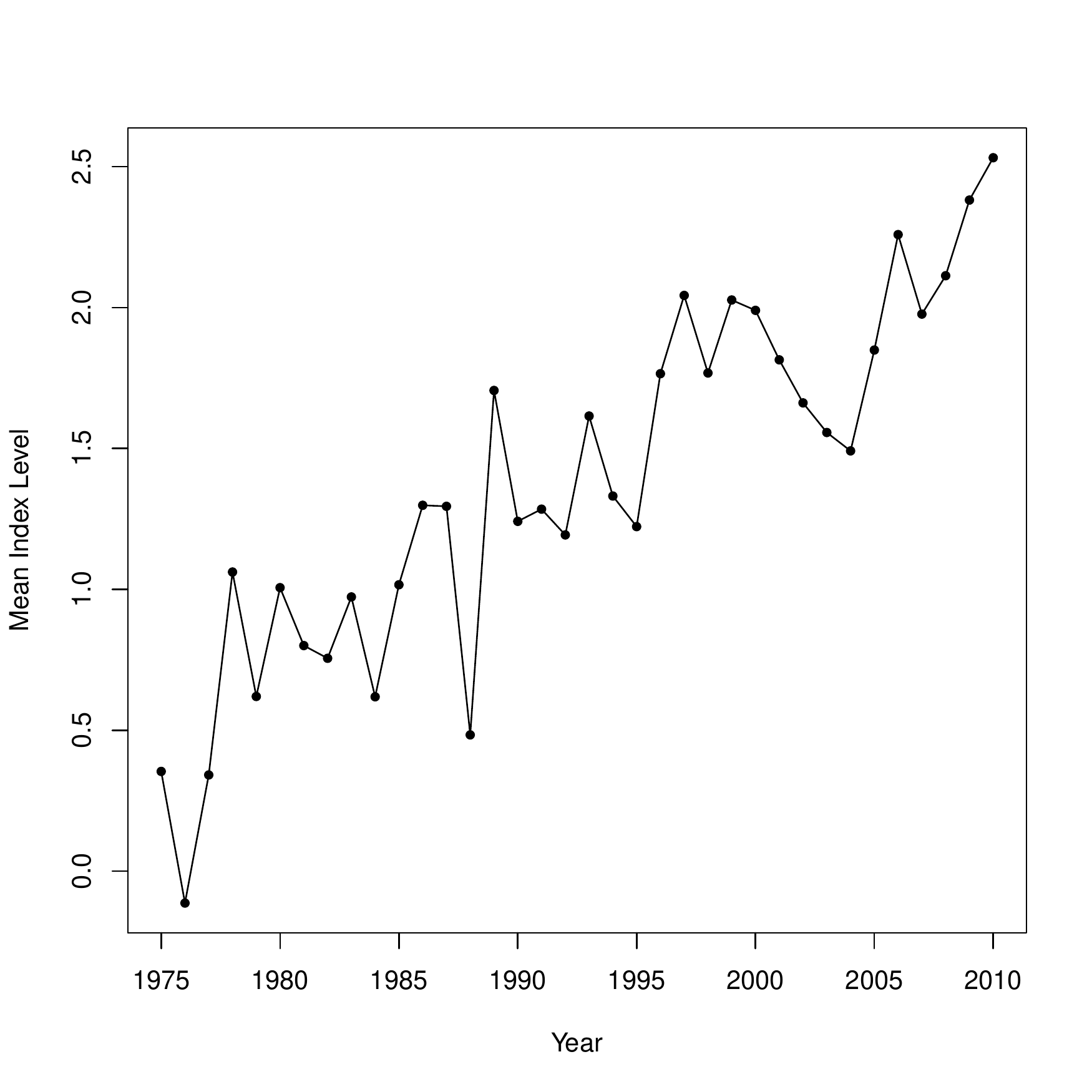}}
  \subfigure[Aggregate]{\includegraphics[width = .32\linewidth]{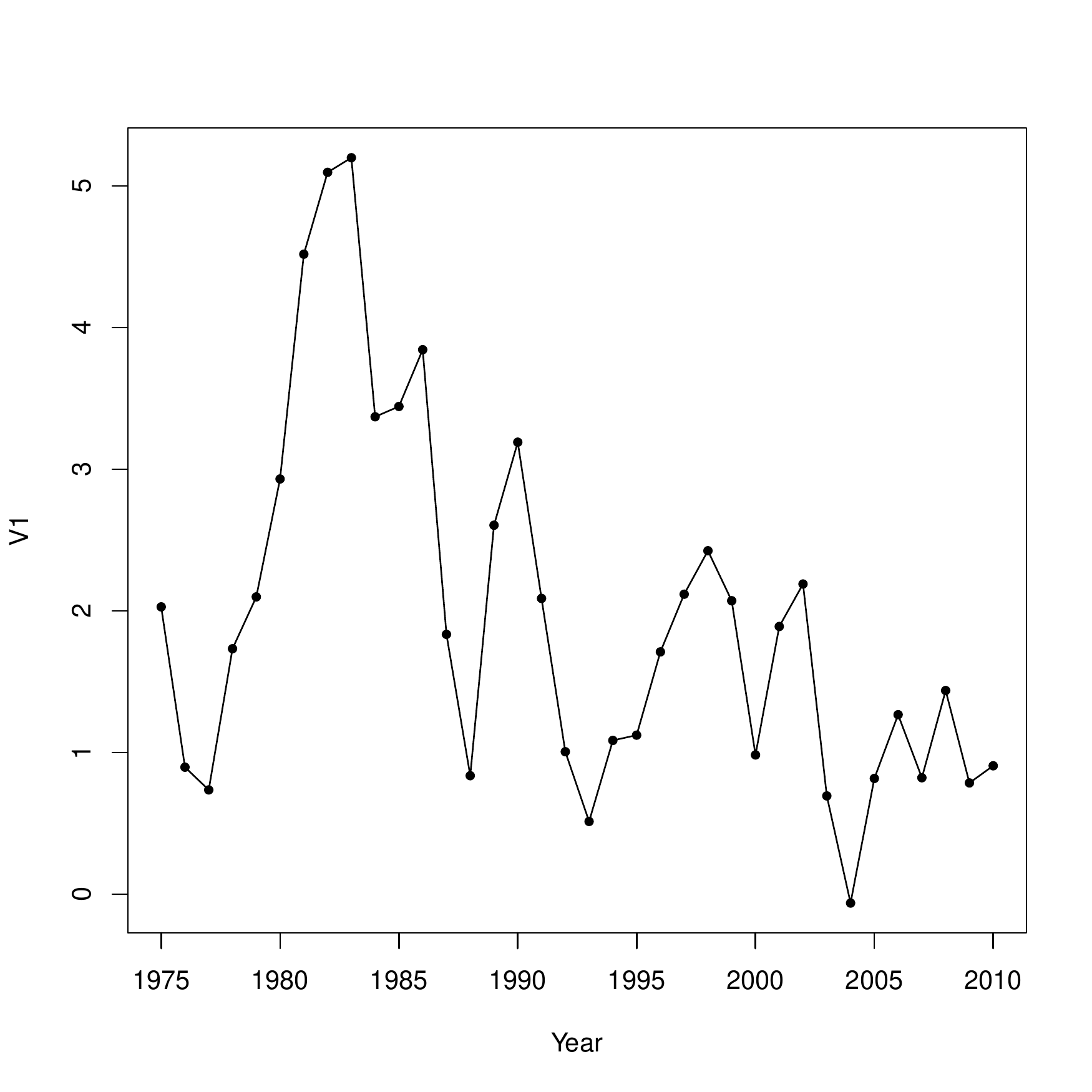}}
  \caption{Mean level of Risk Indices by Year}\label{fig:means}
\end{figure}

\indent Figure~\ref{fig:means} shows the average across all countries for each index by year.  A few things become apparent from this figure.  First, all indices except the Political index appear to achieve their highest levels in the early to mid-eighties, a high-interest rate period with substantial propensity for default, inflation and high unemployment.  Indeed, all four of these risk indices, on the aggregate appear to decline as we move towards 2010.  The result in Figure~\ref{fig:means} (e), which shows a consistently rising mean Political risk index would at first seem quite compelling, implying some increasing risk due to political factors.  Unfortunately, this mainly exposes a failing of the History proxy.  This proxy measures the total number of defaults a country experienced in the past and is therefore consistently increasing.  See our discussion in Section~\ref{sec:Conclude} for potential avenues to create a more robust set of Political proxies.\\
\indent Finally, we address an issue related to theory index portability.  In the theoretical construction of these indices we specified an independence structure between indices.  However, there has been nothing enforcing this condition in posterior estimation.  If theory indices were correlated in the posterior, this would be acceptable, however it would imply that these indices would need to included as a set when attempting to model other phenomena.  Table~\ref{tab:index_correlation} suggests such considerations are likely unnecessary.  In Table~\ref{tab:index_correlation} we show the posterior correlation matrix of the theory indices, estimated over all samples and country/year pairs.  We see in general a low degree of correlation (the entry .-148 between the Illiquidity and Political theories being the highest in absolute value).  This feature is desirable, since it suggests that the theory indices can be used on an individual basis for subsequent modeling of other issues related to economic collapse.
\begin{table}\caption{Posterior correlation matrix of theory indices.  This table mainly shows that the indices have the desirably property of low dependence between one another.}\label{tab:index_correlation}
  \begin{center}
    \begin{tabular}{llllll}
\hline\hline
  & Insolvency & Illiquidity & Macroeconomic & Political & Systemic\\
\hline
Insolvency & 1 & 0.016 & 0.058 & -0.036 & 0.019\\
Illiquidity & 0.016 & 1 & -0.04 & -0.148 & -0.03\\
Macroeconomic & 0.058 & -0.04 & 1 & 0.005 & 0.006\\
Political & -0.036 & -0.148 & 0.005 & 1 & -0.007\\
Systemic & 0.019 & -0.03 & 0.006 & -0.007 & 1\\
\hline\hline
    \end{tabular}
  \end{center}
\end{table}

\section{Conclusions}\label{sec:Conclude}
\indent We have constructed a system whose purpose is to create indices representing various theories which are believed to drive heterogeneity in economic outcomes.  When constructing an index, interpretability
is an important feature to retain.  This is primarily because through interpretability additional proxies can be found when deficiencies become apparent, and specific results can be explained directly.  Our BTA approach
then forms a natural means of incorporating and resolving the obvious model uncertainty present in such a specification.  Furthermore, our focus on modeling multiple outcomes coupled with the ability to entertain a broad set of
outcome sampling distributions lends our system both generalizability and flexibility.\\
\indent There is considerable additional work to be done, both on the technical, algorithmic sides of BTA and also related to the specific goal of modeling an economy's potential for collapse.
One key point has been the assumption that the multiple outcome variables are independent from one another.  In practice, this did not seem to be overly critical, as seen by the fact that the inflation outcome was not present in the posterior when BTA
was run on default using this feature alongside the others shown above.  However, incorporating outcome variable dependence should be relatively straightforward using the Gaussian copula approach of \citet{hoff_2007}.  Indeed uncertainty over these conditional independence assumptions could also be model averaged using the copula Gaussian graphical model approach of \cite{dobra_lenkoski_2011}.\\
\indent In this current system, outcome equations had a linear dependence on theory indices.  While it will always be necessary to orientate the indices for reasons of identification (i.e. the assumption that $\gamma_{rt} = 1$ for at least one non-zero $r$), expanded linear forms such as spline models (\cite{wood2017generalized}) are entirely feasible.  Indeed a third layer of model selection would be to test between linearity and the expanded linearity offered by spline modeling.\\
\indent The MCMC algorithm necessary to resolve BTA was neither trivial nor the most complex.  As outlined as early as \citet{rue_2001}, block updates of parameters in hierarchical generalized models is often advantageous.  We have in general avoided block updates at present, but such a sampling regime could speed up convergence and also algorithm run-time.\\
\indent One difficultly we experience when implementing the quantile regression was the null second derivative in the asymmetric Laplace distribution.  This in turn, makes intelligent updates of parameters for this distribution somewhat harder, since there is less information regarding posterior curvature and thus proposals have a tendency to move too far along the posterior density surface.  This feature has already been investigated in some detail in related contexts.  One potential for improved mixing would be to follow \citet{fasiolo2017fast}, who propose a smooth version of the pinball loss to aid the fitting of qgam models.\\
\indent Finally, our reversible jump proposals were in some sense the least inspired part of the current system.  Though mixing appeared acceptable, more focused jumps could have been constructed, by following much the same Laplacian formulations as the other model parameters.\\
\indent As the great expansionary period following the global financial crisis enters its second decade, it is clear that we can expect to enter a retractionary phase of the global business cycle sometime in the near future.  Our applied interest has been to begin building a monitoring, forecasting and inferential toolset that can prepare us for this period.  While we believe the current version of the SRI estimation system is encouraging, considerable work remains to be done.\\
\indent First and foremost, the current dataset is available until 2010.  We intend to continue building this system to include all available years to present. We are broadly happy with the proxies collected to model insolvency and illiquidity in an economy. Macroeconomic and Systemic features could likely be expanded in a number of obvious ways.  For instance including information on global financial markets or personal or industrial bankruptcy information could expand the Systemic theory proxies.\\
\indent However, we are convinced that the Political risk proxies can be expanded in several important manners.  First, the History proxy is a useful concept, since it captures the propensity of a given country to consistently default on debt (i.e. Argentina), however as pointed out in Figure~\ref{fig:means} its current construction needs to be adjusted to account for the fact that it is presently monotonic in time.  Secondly, aspects related to political regimes are likely to affect potential for economic collapse.  Merging our data with the regime change dataset of \citet{reich_2002} could be one avenue to account for the effect of differing regimes and overall regime uncertainty.\\
\indent Finally, it has been our hope to use only publicly available data sources to aid in the reproducability of our index construction.  While we are convinced that devaluation matters should be included in our set of outcome equations, the necessary currency data has been hard to find publicly.  We will continue to investigate open and public sources of currency exchange data to increase the coverage of this variable.  In doing so, we hope the relative inconclusivity related to theories and their effect on sudden devaluations can be resolved.
  \bibliographystyle{apalike}
  \bibstyle{apalike}
\bibliography{Bibo}

\begin{thebibliography}{}

\bibitem[Brock et~al., 2003]{brock2003policy}
Brock, W.~A., Durlauf, S.~N., and West, K.~D. (2003).
\newblock Policy evaluation in uncertain economic environments.
\newblock Technical report, National Bureau of Economic Research.

\bibitem[Chen et~al., 2017]{chen2017determinants}
Chen, R.-B., Chen, Y.-C., Chu, C.-H., and Lee, K.-J. (2017).
\newblock On the determinants of the 2008 financial crisis: A {B}ayesian
  approach to the selection of groups and variables.
\newblock {\em Studies in Nonlinear Dynamics \& Econometrics}, 21(5).

\bibitem[Dobra and Lenkoski, 2011]{dobra_lenkoski_2011}
Dobra, A. and Lenkoski, A. (2011).
\newblock Copula {G}aussian graphical models and their application to modeling
  functional disability data.
\newblock {\em The Annals of Applied Statistics}, 5(2A):969--993.

\bibitem[Durlauf et~al., 2012]{durlauf_et_2012}
Durlauf, S.~N., Kourtellos, A., and Tan, C.~M. (2012).
\newblock Is god in the details? a reexamination of the role of religion in
  economic growth.
\newblock {\em Journal of Applied Econometrics}, 27(7):1059--1075.

\bibitem[Dyrrdal et~al., 2015]{dyrrdal_et_2015}
Dyrrdal, A.~V., Lenkoski, A., Thorarinsdottir, T.~L., and Stordal, F. (2015).
\newblock Bayesian hierarchical modeling of extreme hourly precipitation in
  norway.
\newblock {\em Environmetrics}, 26(2):89--106.

\bibitem[Fasiolo et~al., 2017]{fasiolo2017fast}
Fasiolo, M., Goude, Y., Nedellec, R., and Wood, S.~N. (2017).
\newblock Fast calibrated additive quantile regression.
\newblock {\em arXiv preprint arXiv:1707.03307}.

\bibitem[Gamerman and Lopes, 2006]{gamerman2006markov}
Gamerman, D. and Lopes, H.~F. (2006).
\newblock {\em {M}arkov chain {M}onte {C}arlo: stochastic simulation for
  {B}ayesian inference}.
\newblock Chapman and Hall/CRC.

\bibitem[Green, 1995]{green_1995}
Green, P.~J. (1995).
\newblock Reversible jump {M}arkov chain {M}onte {C}arlo computation and
  {B}ayesian model determination.
\newblock {\em Biometrika}, 82(4):711--732.

\bibitem[Hastie and Tibshirani, 1990]{hastie2017generalized}
Hastie, T. and Tibshirani, R. (1990).
\newblock {\em Generalized additive models}.
\newblock Chapman and Hall/CRC.

\bibitem[Hoff, 2007]{hoff_2007}
Hoff, P.~D. (2007).
\newblock Extending the rank likelihood for semiparametric copula estimation.
\newblock {\em The Annals of Applied Statistics}, 1(1):265--283.

\bibitem[Karl and Lenkoski, 2012]{karl2012instrumental}
Karl, A. and Lenkoski, A. (2012).
\newblock Instrumental variable {B}ayesian model averaging via conditional
  {B}ayes factors.
\newblock {\em arXiv preprint arXiv:1202.5846}.

\bibitem[Kourtellos et~al., 2019]{kourtellos_et_2019}
Kourtellos, A., Lenkoski, A., and Petrou, K. (2019).
\newblock Measuring the strength of the theories of government size.
\newblock {\em to appear, Empirical Economics}, pages 1--38.

\bibitem[Lenkoski, 2013]{lenkoski_2013}
Lenkoski, A. (2013).
\newblock A direct sampler for {G}-{W}ishart variates.
\newblock {\em Stat}, 2(1):119--128.

\bibitem[Ley and Steel, 2009]{ley_steel_2009}
Ley, E. and Steel, M.~F. (2009).
\newblock On the effect of prior assumptions in bayesian model averaging with
  applications to growth regression.
\newblock {\em Journal of applied econometrics}, 24(4):651--674.

\bibitem[Reich, 2002]{reich_2002}
Reich, G. (2002).
\newblock Categorizing political regimes: New data for old problems.
\newblock {\em Democratization}, 9(4):1--24.

\bibitem[Robert and Casella, 2013]{robert2013monte}
Robert, C. and Casella, G. (2013).
\newblock {\em Monte Carlo statistical methods}.
\newblock Springer Science \& Business Media.

\bibitem[Roubini and Manasse, 2005]{Roubini2005}
Roubini, N. and Manasse, P. (2005).
\newblock “rules of thumb” for sovereign debt crises.
\newblock Working paper no. 05/42, IMF.

\bibitem[Rue, 2001]{rue_2001}
Rue, H. (2001).
\newblock Fast sampling of {G}aussian {M}arkov random fields.
\newblock {\em Journal of the Royal Statistical Society: Series B (Statistical
  Methodology)}, 63(2):325--338.

\bibitem[Rue and Held, 2005]{rue_held_2005}
Rue, H. and Held, L. (2005).
\newblock {\em Gaussian Markov random fields: theory and applications}.
\newblock Chapman and Hall/CRC.

\bibitem[Savona and Vezzoli, 2015]{Savona2012}
Savona, R. and Vezzoli, M. (2015).
\newblock Fitting and forecasting sovereign defaults using multiple risk
  signals.
\newblock {\em Oxford Bulletin of Economics and Statistics}, 77(1):66--92.

\bibitem[Steel, 2019]{steel2017model}
Steel, M.~F. (2019).
\newblock Model averaging and its use in economics.
\newblock {\em to appear, Journal of Economic Literature}.

\bibitem[Wood, 2017]{wood2017generalized}
Wood, S.~N. (2017).
\newblock {\em Generalized additive models: an introduction with R}.
\newblock Chapman and Hall/CRC.

\end{thebibliography}
\appendix

\section{Full Algorithm Details}
Based on data $\mc{D}$ we use MCMC to obtain a sample $\{\varsigma^{[1]}, \dots, \varsigma^{[S]}\}$ of the posterior distribution, where each $\varsigma^{[i]}$ contains
\begin{itemize}
\item $M_1, \dots, M_T$ the models associated with theories $1$ through $T$
\item $\bs{\beta}_1, \dots,\bs{\beta}_T$, the coefficient vectors associated with each theory.  Note that by construction $\beta_{jt} = 0$ when $j \not\in M_t$
\item $\bs{\gamma}_1, \dots, \bs{\gamma}_R$ the theory-scaling vectors for each outcome equation $r$.  A $\gamma_{tr}$ can be set to zero, indicating that theory-$t$ is not currently relevant for outcome equation $r$.  For purposes of identification if multiple $\gamma_{tr}$ are non-zero for a given $t$, we set $\gamma_{tr} = 1$ for whichever $r$ is smallest.
\item $\bs{I}_1, \dots, \bs{I}_T$ the latent theory index vectors (each of length $n$) where $I_{it}$ is the current state of the theory $t$ index for observation $i$.  By convention if $\gamma_{tr} = 0$ for all $r$ then $I_{it} = 0$ for all $i$.
\item Global parameters $\theta_{qr}$ in the $R$ outcome equations
\end{itemize}
When moving from $\varsigma^{[s]}$ to $\varsigma^{[s + 1]}$ we utilize four different MCMC strategies, all of which are now relatively standard in the MCMC literature.  These are
\begin{itemize}
\item Gibbs sampling, relevant for updating $\bs{\beta}_t$
\item Conditional Bayes Factors, which are used to update the theory-level models $M_t$
\item Metropolis-Hastings via Laplacian calculations of the log posterior density which are used, in turn, to update theory indices $\bs{I}_t$, global parameters $\theta_{qr}$ and those theory-scaling parameters $\gamma_{tr}$ which are neither constrained to zero or one.
\item Reversible Jump Methods for alternating $\gamma_{tr}$ between being 0 or in $\mathbb{R}$.  Note that the moves here become especially detailed--though primarily in the sense of bookkeeping--when $\gamma_{tr}$ is currently set to $1$, or if $\gamma_{tr}$ is currently zero and $r$ is smaller than all other non-zero $\gamma_{tr'}$.  Finally, this becomes a joint reversible jump move when the model move will either turn-on or shut-off the theory entirely, as both $\gamma_{tr}$ and $\bs{I}_r$ will be affected.
\end{itemize}
The sections below detail each of these approaches individually
\subsection{Gibbs Sampling Update of $\bs{\beta}_t$}
To resample $\bs{\beta}_t$ we note that its posterior distribution
\begin{align*}
  pr(\bs{\beta}_{t} | \cdot) &= pr(\bs{\beta_t}|M_t, \bs{I}_t, \bs{X}_t)\\
  &= pr(\bs{\beta}_{M_t} | \bs{I}_t, \bs{X}_{M_t})
\end{align*}
Where $\bs{\beta}_{M_t}$ and $\bs{X}_{M_t}$ indicate the restriction to those elements and columns of $\bs{\beta}_t$ and $\bs{X}_t$, respectively, associated with the variables in model $M_t$.  We then have that
\begin{align*}
  pr(\bs{\beta}_{M_t} | \bs{I}_t, \bs{X}_{M_t}) \propto pr(I_{t}|\bs{\beta}_{M_t},\bs{X}_{M_t} )pr(\bs{\beta}_{M_t}).
\end{align*}
using standard results of Bayesian regression (see e.g. Hoff 2009), we therefore have that
\begin{align*}
  pr(\bs{\beta}_{M_t} | \bs{I}_t, \bs{X}_{M_t}) &= \mc{N}(\hat{\bs{\beta}}_{M_t}, \bs{\Xi}_{M_t}^{-1})\\
  \bs{\Xi}_{M_t} &= \bs{X}_{M_t}'\bs{X}_{M_t} + \mathbb{I}_{p_{M_t}}\\
  \hat{\bs{\beta}}_{M_t} &= \bs{I}_t\bs{X}_{M_{t}}'\bs{\Xi}_{M_t}^{-1}.
\end{align*}
We therefore resample $\bs{\beta}_{M_t}$ from the multivariate-Normal distribution above.

\subsection{Conditional Bayes Factors to update $M_t$}
Conditional Bayes Factors compare integrated likelihoods for models $M_t$ and a new proposal model $M'_t$, conditioning on the latent indices $\bs{I}_{t}$.  This conditioning then separates the Gaussian regression components on which the models operate from the larger non-Gaussian components in the response equations, leading to an efficient sampling regime.  This efficiency is present both in the availability of closed form calculations to compare models and the relative parsimony of the approach's exposition.\\
\indent In particular, note that
$$
pr(M_t|\mc{D},\cdot) \propto pr(\bs{I}_{t} | M_t) pr(M_t)
$$
which implies that the latent theory indices $\bs{I}_t$ separate the conditional posterior of the model $M_t$ from the data $\mc{D}$ and the associated non-integrable likelihoods.  This term can the be represented by
$$
pr(\bs{I}_{t} | M_t) pr(M_t) = \int_{\bs{\beta}_{M_r}} pr(\bs{I}_t | \bs{\beta}_{M_r}) pr(\bs{\beta}_{M_r}|M_r)d\bs{\beta}_{M_r} pr(M_t)
$$
The integrand above is then 
$$
\int_{\bs{\beta}_{M_r}} pr(\bs{I}_t | \bs{\beta}_{M_r}) pr(\bs{\beta}_{M_r}|M_r)d\bs{\beta}_{M_r} \propto |\bs{\Xi}_{M_t}|^{1/2}\exp\left(\frac{1}{2}\hat{\bs{\beta}}_{M_t}'\bs{\Xi}_{M_t}\hat{\bs{\beta}}_{M_t}\right)
$$
where $\hat{\bs{\beta}}_{M_t}$ and $\bs{\Xi}_{M_t}$ are defined as above.  Similar to the classic MC3 algorithm, models $M_t$ and $M'_t$ are compared via Metropolis-Hastings.
\subsection{Metropolis-Hastings Updates via Laplacian Expansions}
\indent The two sections above dealt with parameters that could effectively be ``conditioned'' away from the sampling model of the dependent variables, in both cases by conditioning on the latent variables $\bs{I}_t$.  This, in turn, led to updates that were straightforward to calculate as in both cases they relied on well-known results for integrals over the Gaussian distribution.  However, when conditional posterior distributions do not have a form amenable to integration or Gibbs sampling, Metropolis-Hastings algorithms provide an obvious alternative.  This section therefore details all proposal distributions and acceptance ratios necessary to update these parameters.\\
\indent In all cases, we follow a standard approach to creating Gaussian proposals which require no pre-specified tuning parameters and instead adapt proposals to the local curvature of the log posterior density, see e.g. chapter 4 of \cite{rue_held_2005} for a detailed discussion of this approach and \cite{dyrrdal_et_2015} for a similar algorithmic design.  More involved methods, such as Hamiltonian MCMC, Manifold MCMC etc which build on these concepts could have been entertained but mixing was already sufficiently acceptable that these more sophisticated methodologies seemed unnecessary.  See our discussion the Section~\ref{sec:Conclude}.
Suppose, in general, that we would like to update a parameter $\tau$ and write $\log pr(\tau|\cdot) = f(\tau)$ to represent the log posterior density of this parameter with respect to the observations and all other parameters. For designing the proposal distribution, we employ a Gaussian approximation of this posterior density. A quadratic Taylor expansion of the log-posterior $f(\tau)$ around the value $\tau$ gives 
\bl\begin{align*}
f(\tau') & \approx f(\tau) + f'(\tau) ( \tau' - \tau) + \frac{1}{2} f''(\tau) (\tau' - \tau)^2 \\
& = a + b \tau' - \frac{1}{2} c (\tau')^2, 
\end{align*}\el
where $b = f'(\tau) - f''(\tau) \tau$ and $c = -f''(\tau)$.  The posterior distribution $pr(\tau | \cdot)$ can therefore be approximated by 
\bl\[
\widetilde{pr}(\tau | \cdot) \propto \exp \Big( -\frac{1}{2} c (\tau')^2 + b \tau' \Big), 
\]\el 
the density of the Gaussian distribution $\mc{N}(b/c, c^{-1})$. Using this relationship, we choose $\mc{N}(b/c, c^{-1})$ as our proposal distribution, where $\tau$ is the current state in the MCMC chain.  This formulation alleviates the user from specifying a large number of sampling tuning parameters and achieves high acceptance proportions.\\
\indent The following subsections outline the specific forms of $f,f',$ and $f''$ for all variates that are updated in this manner.  Since the $I_{it}$ depend on all $r$ equations they are handled in a final, separate subsection.
\subsubsection{Logistic Regression}
If equation $r$ is a logistic model then it has the form
$$
pr(Y_{ir}|\cdot) = \left(\frac{\exp(\mu_{ir})}{1 + \exp(\mu_{ir})}\right)^{Y_{ir}}\left(\frac{1}{1 + \exp(\mu_{ir})}\right)^{1 - Y_{ir}}
$$
where
$$
\mu_{ir} = \alpha_r + \sum_{t =1}^T \gamma_{rt} I_{it}
$$
And thus the formulas for $\alpha_r$, $\gamma_{rt}$ require derivation (as noted above we leave $I_{it}$ to a final subsection).  First, note
$$
\log pr(Y_{ir}) = Y_{ir}\mu_{ir} - \log(1 + \exp(\mu_{ir}))
$$
Then for the global parameter $\alpha_r$ with prior distribution $\alpha_r \sim \mc{N}(0,1)$ we have that
\begin{align*}
  f(\alpha_r) &= \sum_{i = 1}^{n} \left\{Y_{ir}\mu_{ir} - \log(1 + \exp(\mu_{ir}))\right\} - \frac{\alpha_r^2}{2}\\
  f'(\alpha_r) &= \sum_{i = 1}^n \left\{Y_{ir} - \frac{\exp(\mu_{ir})}{(1 + \exp(\mu_{ir}))}\right\} - \alpha_r\\
  f''(\alpha_r) &= -\sum_{i = 1}^{n} \left\{\frac{\mu_{ir}}{(1 + \mu_{ir})^2}\right\} - 1
\end{align*}
Similarly, for $\gamma_{rt}$ not constrained to be $0$ or $1$ we assume $\gamma_{rt} \sim \mc{N}(0,1)$ and have 
\begin{align*}
  f(\gamma_{rt}) &= \sum_{i = 1}^{n} \left\{Y_{ir}\mu_{ir} - \log(1 + \exp(\mu_{ir}))\right\} - \frac{\gamma_{rt}^2}{2}\\
  f'(\gamma_{rt}) &= \sum_{i = 1}^n \left\{Y_{ir}I_{it} - I_{it}\frac{\exp(\mu_{ir})}{(1 + \exp(\mu_{ir}))}\right\} - I_{it}\\
  f''(\gamma_{rt}) &= -\sum_{i = 1}^{n} I_{it}^2\left\{\frac{\exp(\mu_{ir})}{(1 + \exp(\mu_{ir}))^2}\right\} - 1
\end{align*}
Finally, as it will be important in derivations for the updates of $I_{it}$ we write
\begin{align*}
  l_r(Y_{ir},I_{it}) &= Y_{ir}\mu_{ir} - \log(1 + \exp(\mu_{ir}))\\
  \dot{l}_r(Y_{ir},I_{it}) &= \gamma_{ir}Y_{ir} - \gamma_{ir}\frac{\exp(\mu_{ir})}{1 + \exp(\mu_{ir})}\\
  \ddot{l}_r(Y_{ir}, I_{it}) &= - \gamma_{ir}^2\frac{\exp(\mu_{ir})}{1 + \exp(\mu_{ir})^2}\\
\end{align*}
\subsubsection{Bayesian Quantile Regression}
  Let
  \begin{align*}
    pr(Y_{ir}|\mu_{ir}, \kappa, q) &\propto \exp\left\{\kappa - e^\kappa\rho_q(Y_{ir} - \mu_{ir})\right\} \\
    \rho_q(Y_{ir} - \mu_{ir}) &= (Y_{ir} - \mu_{ir})(q - \mathbf{1}\{Y_{ir} < \mu_{ir}\})
  \end{align*}
  be a Bayesian Quantile Regression, i.e. $Y_{ir}$ is considered asymmetric Laplace distributed with log-precision parameter $\kappa$ and
  \begin{align*}
    \mu_{ir} &= \alpha_{ir} + \sum_{t = 1}^{T} \gamma_{rt}I_{it}.
  \end{align*}
  We therefore need to derive the relevant formulas for $\alpha_{ir}$, $\gamma_{rt}$ and likelihood derivatives for $I_{it}$.  We note
  \begin{align*}
    \log pr(Y_{ir} |  \mu_{ir}, \kappa, q) &= \kappa - e^{\kappa}\rho_q(Y_{ir} - \mu_{ir})
  \end{align*}
  and thus,
  \begin{align*}
    \frac{\partial \log pr(Y_i | \cdot)}{\partial \mu_i} &= e^{\kappa}(q - \mathbf{1}\{Y_i < \mu_i\})\\
    \frac{\partial^2 \log pr(Y_i | \cdot)}{(\partial \mu_i)^2} &= 0.
  \end{align*}
  Therefore, for $\alpha_r$ with $\mc{N}(0,1)$ prior we have
  \begin{align*}
    f(\alpha_r) &= \kappa - e^{\kappa}\sum_{i = 1}^n \rho_q(Y_{ir} - \mu_{ir}) - \frac{\alpha_r^2}{2}\\
    f'(\alpha_r) &= e^{\kappa}\sum_{i = 1}^n(q - \mathbf{1}\{Y_i < \mu_i\}) - \alpha_r\\
    f''(\alpha_r) &= -1.
  \end{align*}
  Similarly when $\gamma_{rt}$ is not constrained to $0$ or $1$ we set $\gamma_{rt} \sim \mc{N}(0,1)$ and have
  \begin{align*}
    f(\gamma_{rt}) &= \sum_{i = 1}^n\left\{\kappa - e^{\kappa}\rho_q(Y_{ir} - \mu_{ir})\right\} - \frac{\gamma_r^2}{2}\\
    f'(\gamma_{rt}) &= \sum_{i = 1}^n \left\{I_{it}e^{\kappa}(q - \mathbf{1}\{Y_i < \mu_i\})\right\} - \gamma_r\\
    f''(\gamma_{rt}) &= -1.
  \end{align*}
  Likewise, we note that
  \begin{align*}
    \frac{\partial \log pr(Y_i | \cdot)}{\partial \kappa} &= 1 - e^{\kappa}(q - \mathbf{1}\{Y_i < \mu_i\})\\
    \frac{\partial^2 \log pr(Y_i | \cdot)}{(\partial \kappa)^2} &= -e^{\kappa}(q - \mathbf{1}\{Y_i < \mu_i\}).
  \end{align*}
  and thus if $\kappa \sim \mc{N}(0,1)$ in the prior, then
  \begin{align*}
    f(\kappa | \cdot) &= n\kappa - e^{\kappa}\sum_{i = 1}^{n} \rho_q(Y_i - \mu_i) - \frac{1}{2}\kappa^2\\
    f'(\kappa | \cdot) &= n - e^{\kappa}\sum_{i = 1}^n \rho_q(Y_i - \mu_i) - \kappa\\
    f''(\kappa | \cdot) &= -e^{\kappa}\sum_{i = 1}^n \rho_q(Y_i - \mu_i) - 1.
  \end{align*}
  Finally, for $I_{it}$ we have 
  \begin{align*}
    l(Y_{ir},I_{it}) &= \kappa - e^{\kappa}\rho_q(Y_{ir} - \mu_{ir})\\
    \dot{l}(Y_{ir}, I_{it}) &= \gamma_{rt} e^{\kappa}(q - \mathbf{1}\{Y_i < \mu_i\})\\
    \ddot{l}(Y_{ir}, I_{it}) &= 0
  \end{align*}
  \subsubsection{GEV Regression}
  When $Y_{ir}$ has the form of a GEV Regression with global log-precision $\kappa$ and shape $\xi$ we have
\begin{align*}
  pr(Y_{ir} | \mu_{ir}, \kappa, \xi) &=  e^{\kappa} h(Y_{ir})^{-(\xi + 1)/\xi} \exp \Big( - h(Y_{ir})^{-\xi^{-1}}\Big)\\
  h(Y_{ir}) &= 1 + \xi e^{\kappa} (Y_{ir} - \mu_{ir})\\
  \mu_{ir} &= \alpha_{r} + \sum_{t = 1}^{T}\gamma_{rt}I_{it}.
\end{align*}
with the additional restriction that $h(\cdot) > 0$.  Calculations for this density have a tendency to become somewhat involved.  We first note
\begin{align*}
a(Y_{ir}) \equiv \log pr(Y_{ir} | \mu_{ir}, \kappa, \xi) &= \kappa - \frac{\xi + 1}{\xi}\log h(Y_{ir}) - h(Y_{ir})^{-\xi^{-1}}
\end{align*}
Since $\partial h(Y_{ir})/\partial\mu_{ir} = -e^{\kappa}\xi$ we have that
\bl\begin{align*}
\dot{a}(Y_{ir}) \equiv \pderiv{}{\mu_{ir}} \log pr(Y_{ir} | \cdot) & = (\xi + 1)e^\kappa h(Y_{ir})^{-1} - e^\kappa h(Y_{ir})^{-\xi^{-1}- 1} \\
\ddot{a}(Y_{ir}) \equiv \pderivsq{\mu_{ir}} \log pr(Y_{ir} | \cdot) &=\xi(\xi + 1)e^{2\kappa}h(Y_{ir})^{-2} - (\xi + 1)e^{2\kappa}h(Y_{ir})^{-\xi^{-1} - 2}.
\end{align*}\el
Therefore, to update $\alpha_r \sim \mc{N}(0,1)$ we have
\begin{align*}
  f(\alpha_r) &= \sum_{i = 1}^n a(Y_{ir}) - \frac{\alpha^2_r}{2}\\
  f'(\alpha_r) &= \sum_{i = 1}^{n}\dot{a}(Y_{ir}) - \alpha\\
  f''(\alpha_r) &= \sum_{i = 1}^n\ddot{a}(Y_{ir}) - 1.
\end{align*}
Likewise, to update any $\gamma_{rt}$ not constrained to $0$ or $1$ we have
\begin{align*}
  f(\gamma_{rt}) &= \sum_{i = 1}^n a(Y_{ir}) - \frac{\gamma_{rt}^2}{2}\\
  f'(\gamma_{rt}) &= \sum_{i = 1}^n \dot{a}(Y_{ir})I_{rt} - \gamma_{rt}\\
  f''(\gamma_{rt}) &= \sum_{i = 1}^n \ddot{a}(Y_{ir})I_{rt}^2 - 1.
\end{align*}
For the term $I_{rt}$ we note
\begin{align*}
  l(Y_{ir}, I_{it}) &= a(Y_{ir})\\
  \dot{l}(Y_{ir}, I_{it}) &= \dot{a}(Y_{ir})\gamma_{rt}\\
  \ddot{l}(Y_{ir}, I_{it}) &= \ddot{a}(Y_{ir})\gamma^2_{rt}.
\end{align*}
Now focus on the global log precision term $\kappa \sim \mc{N}(0,1)$ we have
\begin{align*}
  f(\kappa) &= \sum_{i = 1}^n a(Y_{ir}) - \frac{\kappa^2}{2}\\
  f'(\kappa) &= \sum_{i = 1}^{n}\left\{1 - e^{\kappa}(\xi + 1)(Y_{ir} - \mu_{ir}) + b_1(Y_{ir})\right\} - \kappa\\
  f''(\kappa) &= \sum_{i = 1}^n\left\{-e^{\kappa}(\xi + 1)(Y_{ir} - \mu_{ir}) + b_1(Y_{ir}) - b_2(Y_{ir})\right\} - 1
\end{align*}
where
\begin{align*}
  b_1 &= e^{\kappa}(Y_{ir} - \mu_{ir})h(Y_{ir})^{-\xi^{-1} - 1}\\
  b_2 &= (\xi + 1)e^{2\kappa}(Y_{ir} - \mu_{ir})^2h(Y_{ir})^{-\xi^{-1} - 2}.
\end{align*}
The calculations for the shape parameter $\xi$ are somewhat more involved. Let
\bl\begin{align*}
g_1(Y_{ir}) &= \frac{\xi + 1}{\xi}\log h(Y_{ir})\\
g_2(Y_{ir}) &= \exp\left\{-(\xi^{-1} + 1) \log h(Y_{ir})\right\}
\end{align*}\el
We then obtain 
\bl\begin{align*}
\dot{g}_1(Y_{ir}) & = \pderiv{g_1(Y_{ir})}{\xi} = -\frac{\log h(y_{ts})}{\xi^2} + \frac{\xi + 1}{\xi}h(Y_{ir})^{-1}e^{\kappa}(Y_{ir} - \mu_{ir})\\
\dot{g}_2(Y_{ir}) & = \pderiv{g_2(Y_{ir})}{\xi} = g_2  \left[\frac{\log h(Y_{ir})}{\xi^2} - (\xi^{-1} + 1)h(Y_{ir})^{-1}e^\kappa(Y_{ir} - \mu_{ir})\right],
\end{align*}\el
from which it follows that 
\bl\[
\pderiv{}{\xi} \log pr(Y_{ir}| \cdot) = -\dot{g}_1 - \dot{g}_2.
\]\el
For the second derivative, similar calculations return
\bl\[
\pderivsq{\xi} \log pr(Y_{ir}|\cdot) = \pderiv{}{\xi} \big(-\dot{g}_1 - \dot{g}_2\big) = d_1 + d_2 - d_3 + d_4,
\]\el
where
\bl\begin{align*}
d_1 &= -2\xi^{-3}\log h(Y_{ir}) + \xi^{-2}h(Y_{ir})^{-1} e^\kappa (Y_{ir} - \mu_{ir})\\
d_2 &= \frac{h(Y_{ir})^{-1}(Y_{ir} - \mu_{ir})e^\kappa}{\xi^2} + \frac{\xi + 1}{\xi}h^{-2}(Y_{ir})(Y_{ir} - \mu_{ir})^2e^{2\kappa} \\
d_3 &= \dot{g}_2(Y_{ir})\left[\frac{\log h(Y_{ir})}{\xi^2}\right] + g_2(Y_i) \left[-\frac{2\log h(Y_{ir})}{\xi^{3}} + \frac{h(Y_{ir})^{-1}e^{\kappa}(Y_{ir} - \mu_{ir})}{\xi^{2}}\right]\\
d_4 &= \dot{g}_2(Y_{ir})\left[\frac{h(Y_{ir})^{-1}e^\kappa(Y_{ir} - \mu_{ir})}{\xi}\right] - g_2(Y_i)(Y_{ir} - \mu_{ir})e^\kappa\left[\frac{h(Y_{ir})^{-1}}{\xi^{2}} + \frac{h(Y_{ir})^{-2}(Y_{ir} - \mu_{ir})e^\kappa}{\xi}\right]. 
\end{align*}\el
Hence, for updating $\xi \sim \mc{N}(0,1)$ we have
\begin{align*}
  f(\xi|\cdot) &= \sum_{i = 1}^n \left\{\kappa - g_1(Y_{ir}) - g_2(Y_{ir})\right\} - \frac{\xi^2}{2}\\
  f'(\xi | \cdot) &= \sum_{i = 1}^n \left\{-\dot{g}_1(Y_{ir}) - \dot{g}_2(Y_{ir})\right\} - \xi\\
  f''(\xi | \cdot) &= \sum_{i = 1}^n \left\{d_1 + d_2 - d_3 + d_4\right\} - 1.
\end{align*}
\subsubsection{Updating Theory Indices}
We now consider updating of theory indices $I_{it}$.  Noting that 
$$
I_{it} = \bs{X}'_{rt}\bs{\beta}_{t} + \epsilon_{it}, \quad \epsilon \sim \mc{N}(0,\nu_t^{-1})
$$
We have the formulas
\begin{align*}
f(I_{it} | \cdot) &= \sum_{r = 1}^R l_r(I_{it} | \cdot) - \frac{\nu_t}{2}(I_{it} - \bs{X}'_{rt}\bs{\beta}_t)^2\\
f'(I_{it} | \cdot) &= \sum_{r = 1}^R \dot{l}_r(I_{it} | \cdot) - \nu_t(I_{it} - \bs{X}'_{rt}\bs{\beta}_t)\\
f''(I_{it} | \cdot) &= \sum_{r = 1}^R \ddot{l}_r(I_{it} | \cdot) - \nu_t
\end{align*}
Were the $l_r,\dot{l}_r$ and $\ddot{l}_r$ terms are those discussed in the sections above for each respective outcome equation $r$ in the system.
\subsection{Updating Theory Inclusion Parameters via Reversible Jump}
Suppose now that $\gamma_{rt} = 0$ in the current state of the chain.  In the relatively straightforward case in which there is a an $r' < r$ for which $\gamma_{rt} = 1$ -- and thus the inclusion of the $\gamma_{rt}$ will not affect identification matters, we may attempt to make $\gamma_{rt}$ non zero by proposing $\gamma'_{rt} \sim \mc{N}(0,1)$.  We thus transition from $(\bs{\gamma}_r,\gamma_{rt})$, where $\bs{\gamma}_{rt} = 0$ to $\bs{\gamma}'_t$ with $(\bs{\gamma}'_{t})_r = \gamma_{rt}$ and $(\bs{\gamma}'_t)_s = (\bs{\gamma}_t)_s$ for all other $s \neq r$, a transformation with Jacobian $1$.  Letting
$$
\mu_{ir} = \alpha_r + \sum_{t = 1}^T\gamma_{rt}I_{it}
$$
and 
$$
\mu'_{ir} = \alpha_r + \sum_{t = 1}^T\gamma'_{rt}I_{it}
$$
Since our prior sets all $\gamma_{rs}\sim \mc{N}(0,1)$, the auxiliary density cancels with the larger prior we thus have that
\begin{align*}
  \log pr(\bs{\gamma}_{r}, \gamma'_{rt}|\cdot ) &\propto \sum_{i = 1}^{n} l_r(Y_{it} | \mu_{it})\\
  \log pr(\bs{\gamma}'_r |\cdot) &\propto \sum_{i = 1}^n l_r(Y_{it} | \mu'_{it})
\end{align*}
where $l_r$ is the associated log-likelihood for equation $r$.  This gives the necessary log densities for comparing $\gamma_{rt} \in \mathbb{R}$ and $\gamma_{rt} = 0$.  See our discussion in the Conclusions section regarding more focused proposals of $\gamma_{rt}$ which could aid in mixing and would also make the expressions above slightly more involved.\\
\indent When $\gamma_{rt} = 0$ and $\gamma_{st} = 1$ for $s > r$, some bookkeeping is necessary to adjust the system.  In particular, we sample $\alpha \sim \mc{N}(0,1)$.  We then create a new vector $\bs{\gamma}_t$ where
\begin{equation*}
  \gamma_{st} =
  \begin{cases}
    1, & \text{if}\ s=r \\
    \alpha \gamma'_{st}, & \text{otherwise}
  \end{cases}
\end{equation*}
And similarly we move from $I_{it}$ to $I'_{it}$ by setting $I'_{it} = I_{it} / \alpha$, $\bs{\beta}'_{t} = \bs{\beta} / \alpha$ and $\nu'_t = \alpha \nu_t$.  We therefore note that while we have changed all $\gamma_{st}$ values and the associated theory indices $I_{it}$, only the likelihood for the dependent variable $r$ is affected and comparisons can then be performed as discussed above.\\
\end{document}